\documentclass[10pt,letterpaper,twocolumn]{article} 

\usepackage{ol2}
\usepackage{amsmath}

\newcommand{\rmi}{\mathrm{i}}
\newcommand{\rmd}{\mathrm{d}}

\begin{document}

\twocolumn[ 

\title{Superoscillation in speckle patterns}

\author{Mark R Dennis}

\address{H H Wills Physics Laboratory, University of Bristol, Tyndall Avenue, Bristol BS8 1TL, UK}

\author{Alasdair C Hamilton and Johannes Courtial}

\address{School of Physics and Astronomy, University of Glasgow, Glasgow G12 8QQ, UK}

\begin{abstract}
Waves are superoscillatory where their local phase gradient exceeds the maximum wavenumber in their Fourier spectrum.
We consider the superoscillatory area fraction of random optical speckle patterns. 
This follows from the joint probability density function of intensity and phase gradient for isotropic gaussian random wave superpositions.
Strikingly, this fraction is 1/3 when all the waves in the two-dimensional superposition have the same wavenumber.
The fraction is 1/5 for a disk spectrum.
Although these superoscillations are weak compared with optical fields with designed superoscillations, they are more stable on paraxial propagation.
\end{abstract} 


]

\noindent
The spatial rate of change of a plane wave is determined by its wavevector, which can be defined as the gradient of its phase.
Its direction and magnitude (wavenumber) are unambiguous.
The phase gradient of more complicated wave fields can be used as a definition of local wavevector, which may be a complicated function of position.
Such is the case in superpositions of plane waves.
The modulus of phase gradient is sometimes smaller than the superposition's maximum wavenumber, and sometimes bigger.
This latter case has drawn much attention: at such places, the phase changes more rapidly than the constituent plane waves, hence the term `superoscillation' \cite{berry:1994b262}.
Superoscillatory waves, which locally vary much faster than their fastest Fourier component, have surprising and counterintuitive properties, and have recently been studied in a variety of systems, particularly signal processing, quantum mechanics and optics \cite{kempf:2000blackholes,bp:2006b388,hzcg:2007focusing,tbwyc:2008simulation}.
Quantum mechanically, they fit into the general notion of weak measurements \cite{ar:2005quantum}, and applications in optical imaging science have been suggested \cite{zheludev:2008what?}.

Our purpose here is to study some simple superoscillatory aspects of two-dimensional random waves, that is, superpositions of plane waves whose direction (in a plane) and phase are independent and uniformly distributed random variables.
Such superpositions are a well-established model for speckle patterns -- scalar waves, either optical \cite{goodman:2007speckle} or acoustic \cite{ebeling:1984statistical}, reflected or refracted from random rough surfaces.
In this case, the two dimensions we consider are those of the plane transverse to the overall propagation, and the waves are superoscillatory in the sense that the transverse phase gradient is larger than the maximum transverse wavenumber.
These superpositions are also used as a model for quantum wavefunctions in two-dimensional chaotic enclosures (`chaotic billiards') \cite{stoeckmann:1999quantumchaos,sisb:2002interior}; in particular, when the system is not time-reversal symmetric (either due to absorption, open channels or a magnetic field), the wavefunction is complex.

A significant fraction of the area of a typical speckle pattern, whose transverse wave spectrum is band-limited, is superoscillatory.
Specifically, a wave $\psi = \rho \exp(\rmi \chi),$ dependent on planar position $\boldsymbol{r} = (x,y),$ with intensity $I = \rho^2$ and phase gradient $\nabla \chi,$ is 
\begin{equation}
   \hbox{superoscillatory where }|\nabla \chi|^2 - k_{\mathrm{max}}^2 > 0,
   \label{eq:superosc}
\end{equation}
where $k_{\mathrm{max}}$ is the maximum wavenumber in the transverse superposition spectrum.
This definition of superoscillation originated in a recent study of the relationship between waves and rays in structured refractive materials \cite{hc:2008metamaterials}.
The purpose of this paper is to examine the areas where Eq.~(1) is satisfied in random optical waves.

Optical vortices -- the nodes of wave fields, where the phase is undefined \cite{nb:1974b34} -- may be thought of as extremes of superoscillation, since the phase gradient diverges as $I \to 0.$
Clearly, the vortices lie in superoscillatory regions.
The configuration of vortices in random waves and speckle patterns has been much studied \cite{berry:1978b67,bzmps:1981dislocations,bd:2000b321,goodman:2007speckle}, and the present work may be thought of as generalizing this to all parts of the wave where Eq.~(\ref{eq:superosc}) is satisfied.

The phase gradient $\nabla \chi$ is also related to the current density $\boldsymbol{J} = \operatorname{Im} \psi^* \nabla \psi = I \nabla \chi$.
Therefore $\boldsymbol{J}$ and $\nabla \chi$ are parallel, but have different lengths; as will be demonstrated, superoscillation tends to occur where $I$ is small, so the fluctuations in the $|\nabla \chi|$ distribution are greater than those for $J = |\boldsymbol{J}|.$

The calculation of the superoscillatory fraction of a random wave superposition will use gaussian statistics, as usual in the study of speckle patterns \cite{goodman:1985statistical}, which applies in the limit of infinitely many independent random plane waves. 
For isotropic random two-dimensional waves, the complex fields $\psi$, $\partial_x \psi$ and $\partial_y \psi$ have independent gaussian probability density functions, with variances $\langle |\psi|^2 \rangle = I_0$ (mean intensity), $\langle |\partial_j \psi|^2 \rangle = 2 I_0 k_2,$ where $j = x,y$ and $k_2$ is the normalized second moment of the power spectrum (which is circularly symmetric in $k$-space), with well-defined $k_{\mathrm{max}}.$ 
A natural measure of the correlation length of the random field is $k_2^{-1/2}.$

We begin with the calculation of the joint probability of intensity and current $P(I,J),$ from its Fourier transform,
\begin{eqnarray}
   P(I,\boldsymbol{J}) &=&  \frac{1}{(2\pi)^3} \int \rmd s \int \rmd^2 \boldsymbol{t} \exp(\rmi s I + \rmi \boldsymbol{t}\cdot\boldsymbol{J}) \label{eq:pij}
 \\
   & & \quad   \times \langle \exp(-\rmi s |\psi|^2  + \tfrac{1}{2} \boldsymbol{t}\cdot( \psi^* \nabla \psi - \psi \nabla \psi^*) ) \rangle \nonumber,
\end{eqnarray}
where $\langle \bullet \rangle$ denotes the gaussian average.
This average is straightforward to calculate: in the exponent, the quadratic forms depending on $s$ and $\boldsymbol{t}$ may be added to those from the gaussian probability density, yielding a complex six-dimensional quadratic form matrix with determinant $(2(1+ \rmi I_0 s) + I_0^2 k_2 |\boldsymbol{t}|^2)^2/I_0^6 k_2^4.$
The gaussian average is the reciprocal square root of this determinant, divided by the square root of the product of the variances, $(I_0^6 k_2^4/4)^{1/2}.$
The integrals in $s$ and $\boldsymbol{t}$ may then be found using straightforward complex integration techniques, and then the (equidistributed) direction of $\boldsymbol{J}$ may be integrated.
The final result is
\begin{equation}
   P(I,J) = \frac{J}{I I_0^2 k_2} \exp\left(-\frac{1}{2I_0}(2I + J^2/I k_2)\right).
   \label{eq:pijfin}
\end{equation}
Integrating over $J$ gives the well-known distribution for intensity, $P(I) = \exp(-I/I_0)/I_0,$ and integrating over $I$ gives the previously derived probability density for $J$ in two dimensions, $P(J) = 2 J K_0(\sqrt{2/k_2} J/I_0)/I_0^2 k_2$ (cf.~\cite{ebeling:1984statistical} Eq. (84), \cite{sisb:2002interior} Eq.~(18)), where $K_0$ is a modified Bessel function.

\begin{figure}
   \centerline{\includegraphics[width=7.5cm]{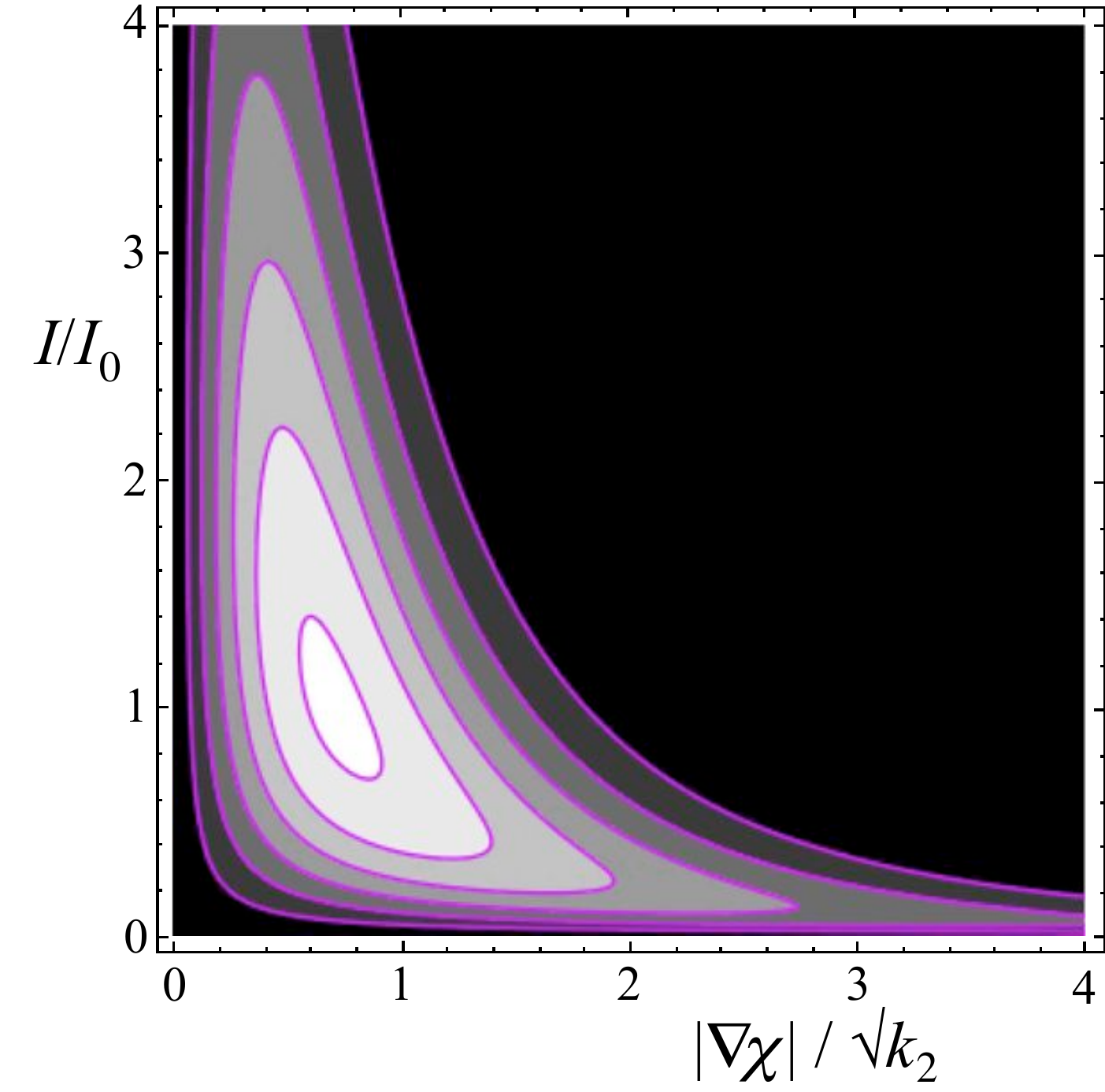}}
   \caption{(Color online) Contour plot of the joint probability density function $P(I,|\nabla \chi|)$ of Eq.~(\ref{eq:pic}). 
   It is clearly unbounded both in $I$ and $|\nabla \chi|,$ although high values of these are anticorrelated. }
   \label{fig:pic}
\end{figure}

Since $J = I |\nabla \chi |,$ the joint probability density function for $I$ and $|\nabla \chi|$ can be found by dividing by the jacobian determinant $|\partial(I,J) / \partial(I,|\nabla \chi|)| = I,$ 
\begin{equation}
   P(I, |\nabla \chi|) = \frac{I |\nabla \chi |}{I_0^2 k_2} \exp\left(-\frac{I}{I_0}(1+|\nabla \chi|^2/2k_2)\right).
   \label{eq:pic}
\end{equation}
This probability distribution green is one of the main results of this Letter, and is plotted in Fig.~\ref{fig:pic}.
It shows quantitatively a clear correlation between low intensities and high phase gradients (and vice versa) in random waves, in line with common wisdom on superoscillation \cite{hc:2008metamaterials}.
Integrating over $I$ gives the probability density function for $|\nabla \chi|$:
\begin{equation}
   P(|\nabla \chi|) = \frac{4 k_2 |\nabla \chi |}{(2k_2 + |\nabla \chi|^2)^2}.
   \label{eq:pc}
\end{equation}
This simple expression was previously derived in \cite{goodman:2007speckle} (Eq.~4-190; also see \cite{og:1983ray} for a more extended discussion). 
The probability distribution of the phase gradient is unbounded, with diverging variance.
Since, for gaussian random waves, area averages are equivalent to ensemble averages \cite{goodman:1985statistical}, the area fraction $f$ of the speckle pattern which is superoscillatory is 
\begin{equation}
   f = \int_{k_{\mathrm{max}}}^{\infty} \rmd |\nabla \chi| \, P(|\nabla \chi|).
   \label{eq:fdef}
\end{equation}

\begin{figure}
   \centerline{\includegraphics[width=9cm]{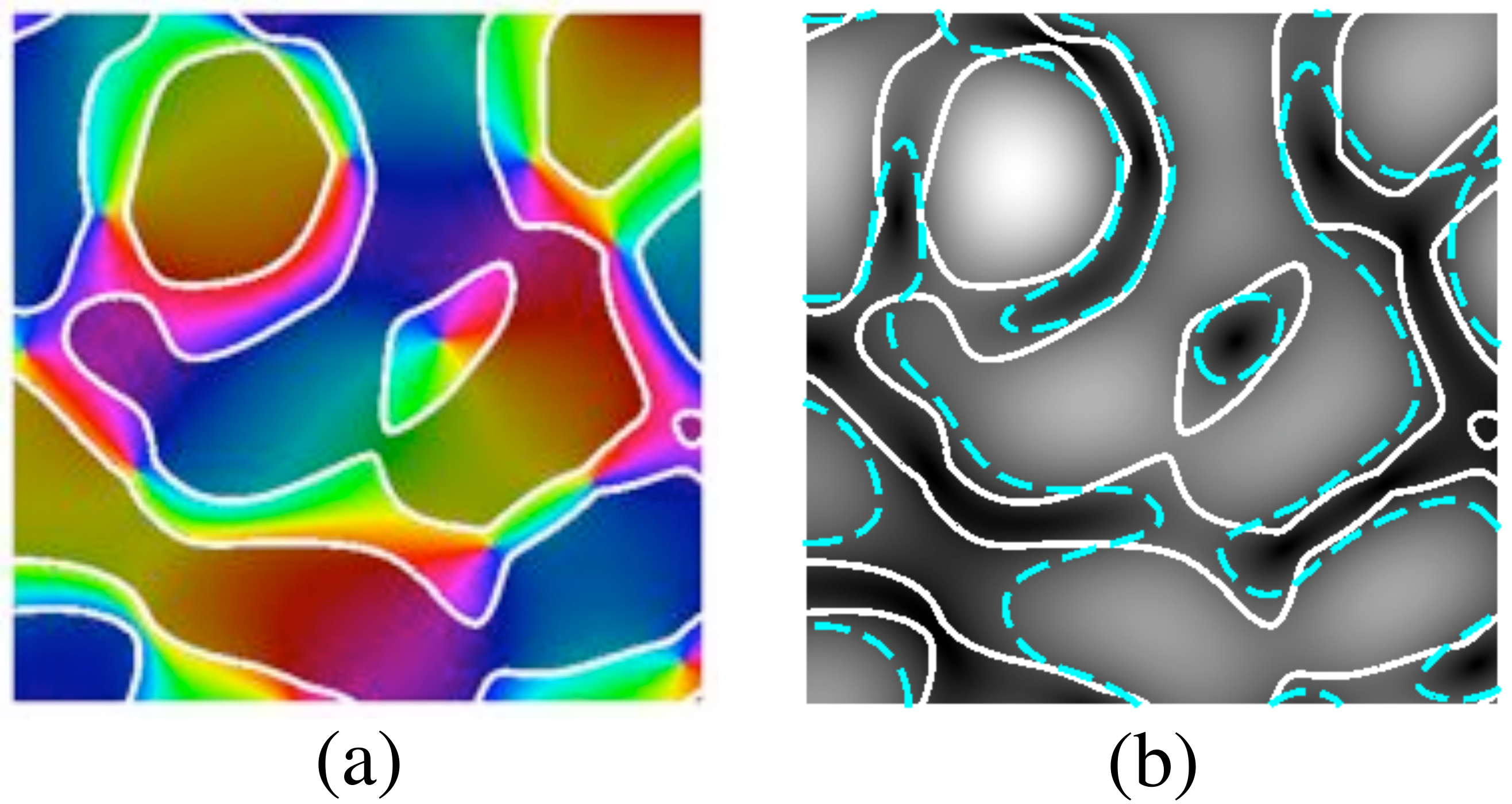}}
   \caption{(Color online) A random superposition of 100 two-dimensional plane waves with the same wavenumber $k.$
   (a) Phase pattern (hues); (b) intensity pattern (grayscale).
   The white contour denotes the line $|\nabla \chi| = k.$  
   The suboscillatory region, occupying statistically 2/3 of the area, is shaded with a dark filter in (a).
   Several phase singularities can be seen in the superoscillatory region of (a).
   The intensity contour (dashed cyan) enclosing the lowest 1/3 of the intensity pattern is also shown in (b).
   This is close to the white superoscillation contour, but the two are clearly different.
   The area plotted is $(4\pi/k)^2.$  }
   \label{fig:superphase}
\end{figure}

The numerical value of the superoscillatory area fraction $f$ depends on the transverse spectrum, specifically the relationship between $k_{\mathrm{max}}$ and $k_2.$
Mathematically, the most natural spectrum to choose is monochromatic waves in the plane.
This corresponds to random wave solutions of the Helmholtz equation $\nabla^2 \psi + k^2 \psi = 0$ ($\nabla^2$ is the two-dimensional laplacian operator).
For paraxial beams, this corresponds to nondiffracting speckle patterns: the spectrum lies on a ring of radius $k$ in transverse Fourier space, and in this case $k_2 = k^2 = k_{\mathrm{max}}^2.$
Putting this into Eq.~(\ref{eq:fdef}) gives $f_{\mathrm{ring}} = 1/3:$ a third of a random monochromatic two-dimensional wave superposition is superoscillatory.
A numerical realization of such a random wave field is shown in Fig.~\ref{fig:superphase}.
Clearly, all of the vortices are in the superoscillatory region.

It is instructive to compare the superoscillatory areas with the area of 1/3-lowest intensity, as in Fig.~\ref{fig:superphase}(b).
Although the areas are similar, the contour $|\nabla \chi| = k$ encloses a slightly different region from the intensity contour, and their topologies are not the same.
Superoscillatory regions are therefore subtly different from low-intensity regions.

It is interesting to observe that, on dividing by $\exp(\rmi \chi),$ the real part of the Helmholtz equation can be rewritten:
\begin{equation}
   |\nabla \chi |^2 - k^2 = \frac{\nabla^2 \rho}{\rho}.
   \label{eq:hhre}
\end{equation}
Therefore, for general monochromatic waves, sub- and superoscillation is governed the laplacian of the wave's real amplitude.
In particular, the superoscillatory boundary contours are given by the nodal lines of $\nabla^2 \rho.$
It is interesting to observe that the right-hand side of Eq.~(\ref{eq:hhre}) is the negative of the `quantum potential' in the hydrodynamic interpretation of quantum mechanics \cite{holland:1993quantum}.

The superoscillatory fraction for other band-limited isotropic spectra is easy to calculate.
For instance, for a top-hat spectrum (disk spectrum) with equal weighting for all waves with wavenumber $k < k_{\mathrm{max}}$, it is easy to show that $k_2 = k_{\mathrm{max}}^2/2$ \cite{goodman:1985statistical}.
The superoscillatory fraction here is then $f_{\mathrm{disk}} = 1/5.$
Interpolating between the two is the case of an annular spectrum, where $k_{\mathrm{max}}(1 - \delta) \le k \le k_{\mathrm{max}},$ where $\delta$ is a scaled thickness: $\delta = 1$ is the disk spectrum, and the limit $\delta \to 0$ is the monochromatic ring spectrum.
For the annular spectrum, $k_2 = k_{\mathrm{max}}^2 (2 - 2\delta + \delta^2)/2,$ and $f_{\mathrm{annulus}} =  1 - 4/(6 - 2 \delta + \delta^2),$ which smoothly interpolates between the two limiting cases.

Of course, several familiar speckle spectra, such as a gaussian distribution of $k$, are not band-limited, and so, strictly speaking, cannot be superoscillatory.
Mathematically, in the paraxial regime -- appropriate for speckle patterns -- all transverse wavenumbers, even infinitely large ones, are infinitesimal compared to the $z$-component of the full wavevector.
The full three-dimensional wavenumber is technically infinite, and there are no true paraxial superoscillations.
However, the transverse wave, considered as a superposition of two-dimensional waves, is superoscillatory as we describe.

It is, of course, possible to study superoscillation in volumes of three-dimensional wave fields, not subject to the drawback of paraxiality (such as the fields relevant to \cite{hc:2008metamaterials}).
A natural choice would be superpositions of isotropically random monochromatic waves in three dimensions, modelling, for example, (scalar) field modes of chaotic cavities \cite{stoeckmann:1999quantumchaos,bd:2000b321}.
These calculations, generalized to $D$ dimensions, appear in a follow-up paper \cite{bd:2008Ddim}.

The stability of transverse, highly superoscillatory fields on paraxial propagation has been previously considered \cite{bp:2006b388}: the superoscillations were found to propagate towards the far field, but ultimately were suppressed by suboscillatory regions.
The naturally-occurring superoscillations considered here have far smaller phase gradients than those in \cite{bp:2006b388}, but their area fraction is statistically constant on propagation.
This observation originates from the fact that, within the paraxial approximation, the power spectrum does not change on propagation.
In the extreme case of transversely monochromatic speckle patterns, the Fourier spectrum itself is invariant on propagation -- the beam is diffraction-free -- and so the superoscillatory regions are themselves invariant on propagation.

We are grateful to Michael Berry, John Hannay, Kevin O'Holleran and Miles Padgett for discussions.
MRD and JC are supported by the Royal Society of London.

\end{document}